\documentclass[12pt]{article}
\usepackage{latexsym}
\usepackage{amsfonts}

\usepackage{amssymb}
\usepackage{epsfig}
\usepackage{amsmath}
\usepackage{amsthm}
\usepackage[mathscr]{eucal}
\usepackage{subfigure}
\usepackage{graphicx}

\newtheorem{Theorem}{Theorem}

\renewcommand{\qed}{\hfill{\ \ \rule{2mm}{2mm}} \vspace{0.2in}}
\newcommand{\ind}{1\hspace{-2.3mm}{1}}

\begin{document}

\title{Linear Codes with Prescribed Hull Dimension and Minimum Distance}
\author{ \textbf{Ghurumuruhan Ganesan}
\thanks{E-Mail: \texttt{gganesan82@gmail.com} } \\
\ \\
IISER Bhopal}
\date{}
\maketitle

\begin{abstract}
The hull of a linear code (i.e., a finite field vector space)~\({\mathcal C}\) is defined to be the vector space formed by the intersection of~\({\mathcal C}\) with its dual~\({\mathcal C}^{\perp}.\) Constructing vector spaces with a specified hull dimension has important applications and it is therefore of interest to study minimum distance properties of such spaces. In this paper, we use the probabilistic method to obtain spaces with a given hull dimension and minimum distance and also derive Gilbert-Varshamov type sufficient conditions for their existence.


\vspace{0.1in} \noindent \textbf{Key words:} Hull dimension, minimum distance, finite field vector spaces, linear codes.

\vspace{0.1in} \noindent \textbf{AMS 2000 Subject Classification:} Primary: 60J35, 15B33;
\end{abstract}

\bigskip

\renewcommand{\theequation}{\arabic{section}.\arabic{equation}}
\renewcommand{\theequation}{\arabic{section}.\arabic{equation}}
\setcounter{equation}{0}
\section{Introduction} \label{intro}
The minimum distance of a vector space or a linear code is a measure of its error correction capability and algebraic constructions are available to obtain codes with specified minimum distance (Huffman and Pless (2010)).  Recently, there has been increasing interest in constructing codes with a given hull dimension, due to wide applications ranging from cryptography to quantum error correction (Carlet et al. (2019)).  The hull of a linear code~\({\mathcal C}\) is simply the intersection of~\({\mathcal C}\) with its dual (formal definitions in Section~\ref{diag_codes}). The hull is itself a linear code and Sendrier (1997) shows that the \emph{expected} hull dimension of a randomly chosen code with a given dimension asymptotically converges to a constant (see also Skersys (2003)). Recently, Sangwisut et al. (2015) and Luo et al. (2018) study linear codes with a given hull dimension and specific parameters, for cyclic linear codes and maximum distance separable codes, respectively and in a related work, Carlet et al. (2019) use prime ideal decompositions to construct codes with a one dimensional hull.



In this paper, we use the probabilistic method to derive sufficient conditions for the existence of linear codes with a \emph{given} hull dimension and minimum distance. In the following Section, we state and prove our main result Theorem~\ref{thm_main}, regarding codes with given hull dimension and minimum distance and illustrate our result with a brief example. For generality, we henceforth use the term vector spaces instead of linear codes, throughout.


\renewcommand{\theequation}{\arabic{section}.\arabic{equation}}
\setcounter{equation}{0}
\section{Hull dimension of vector spaces}\label{diag_codes}
Let~\(q\) be a power of a prime number and let~\(\mathbb{F}_q\) be the finite field containing~\(q\) elements. For integer~\(m \geq 1\) we say that a set of vectors~\({\mathcal G} = \{h_1,\ldots,h_k\} \subset \mathbb{F}_q^{m}\) is a \emph{basis} if the vectors in~\({\mathcal G}\) are linearly independent. If~\({\mathcal G}\) is a basis and~\({\mathcal C} \subset \mathbb{F}_q^{m}\) is the space spanned by~\({\mathcal G},\) then the dimension of~\({\mathcal C}\) is~\(k.\) We then denote~\({\mathcal C}\) to be a~\([m,k]_q-\)space.

We say that a vector~\(x =(x_1,\ldots,x_m) \in \mathbb{F}_q^{m}\) is \emph{orthogonal} to\\\(y = (y_1,\ldots,y_m)\) if~\(x \cdot y^{T} = \sum_{i=1}^{m} x_i \cdot y_i = 0.\) Throughout,~\(T\) in the superscript refers to the transpose operation and all vectors are row vectors. The dual space~\({\mathcal C}^{\perp}\) of a space~\({\mathcal C} \subset \mathbb{F}_q^{m}\) is the set of all vectors~\(v \in \mathbb{F}_q^{m}\) such that~\(v\) is orthogonal to each vector in~\({\mathcal C}.\) The \emph{hull} of~\({\mathcal C}\) is the vector space~\({\mathcal C} \cap {\mathcal C}^{\perp}.\) If~\(k\) and~\(t\) denote the dimension of~\({\mathcal C}\) and its hull respectively, then we say that~\({\mathcal C}\) is a~\([m,k]_q-\)space with hull dimenions~\(t.\) Spaces with hull dimension~\(t=0\) and~\(t=k\) are called complementary dual (Massey (1992)) and self-orthogonal (Kohnert and  Wassermann (2009)) spaces, respectively.

The distance between two vectors~\(x=(x_1,\ldots,x_m)\) and~\(y=(y_1,\ldots,y_m)\) in a space~\({\mathcal C}\) is defined as~\(d(x,y) := \sum_{i=1}^{n} \ind(x_i \neq y_i),\)
where~\(\ind(.)\) refers to the indicator function. The weight of~\(x\) is the number of non-zero entries in~\(x.\) We say that~\({\mathcal C}\) has a minimum distance of at least~\(d\) if any two vectors in~\({\mathcal C}\) have a distance of at least~\(d.\) Because~\({\mathcal C}\) is vector space, this is equivalent to saying that the weight of any vector in~\({\mathcal C}\) is at least~\(d.\)

The following is the main result of the paper.
\begin{Theorem}\label{thm_main} Let~\(m\geq k \geq 1, 1\leq d \leq m\) be integers and let~\(q \geq 2\) be a power of prime satisfying
\begin{equation}\label{q_cond}
1+\sum_{j=0}^{d-1} (q-1)^{j+1} \cdot {m \choose j} < q^{m-2k+2},
\end{equation}
strictly. Letting~\(0 \leq t \leq k\) be any integer we have the following:\\
\((i)\) If~\(q\) is even then there exists an~\([m+k,k]_q-\)space~\({\mathcal C}_1\) with hull dimension~\(t\) and minimum distance at least~\(d.\)\\
\((ii)\) If \(q \equiv 1 \mod{4}\) then there exists a~\([2m+k,k]_q-\)space~\({\mathcal C}_2\) with hull dimension~\(t\) and minimum distance at least~\(2d.\)\\
\((iii)\) If~\(q \equiv 3 \mod{4}\) then there exists a~\([3m+k,k]_q-\)space~\({\mathcal C}_3\) with hull dimension~\(t\) and minimum distance at least~\(3d.\)
\end{Theorem}
The inequality~(\ref{q_cond}) is a sufficient condition that guarantees existence of spaces with a given hull dimension and minimum distance.



Before proving Theorem~\ref{thm_main}, we illustrate~(\ref{q_cond}) with an example. For~\(d-1 \leq \frac{m}{2}\) we use the unimodality of the Binomial coefficient to upper bound the left side of~(\ref{q_cond}) as \[(d+1) \cdot (q-1)^{d} \cdot {m \choose d-1} \leq (d+1) \cdot q^{d} \cdot {m \choose d-1}\]
so that
\begin{equation}\label{q_cond_inf}
(d+1) \cdot {m \choose d-1} < q^{m-2k-d+2}
\end{equation}
is sufficient for~(\ref{q_cond}) to hold. Now let~\(d = \delta \cdot m,t = \gamma \cdot m\) and~\(k = \epsilon \cdot m\) where~\(0 < \gamma<\epsilon<1\) and~\(0 < \delta < \frac{1}{2}\) are constants not depending on~\(m.\) Using the Stirling approximation, we see that the term~\({m \choose d-1}\) grows roughly as~\(2^{mH(\delta)},\)
where~\[H(\delta) := -\delta \cdot \log{\delta} - (1-\delta) \cdot \log(1-\delta)\] is the entropy function and logarithm is to the base~\(2,\)
while~\(q^{m-2k-d+2}\) grows as~\(q^{m(1-2\epsilon-\delta)}.\) Thus if~\(2^{H(\delta)}< q^{1-2\epsilon-\delta}\) or equivalently~\(\epsilon < \epsilon_0(\delta,q) := \frac{1}{2}\left(1-\delta -\frac{H(\delta)}{\log{q}}\right) \) strictly and~\(q\) is even, then there exists a~\([m+k,k]_q-\)space~\({\mathcal C}_1\) with hull dimension~\(t\) and minimum distance at least~\(d.\)

\emph{Proof of Theorem~\ref{thm_main}}:  A set of vectors~\(\{h_1,\ldots,h_k\} \subset \mathbb{F}_q^{m}\) is said to be \emph{mutually orthogonal} if~\[h_i \cdot h_j^{T} = 0 \text{ for any }1 \leq i \neq j \leq k.\]  Our proof consists of two steps: In the first step, we use the probabilistic method to construct a set of mutually orthogonal vectors. Next, we use these mutually orthogonal vectors to construct self-orthogonal vectors and obtain vector spaces with given hull dimension.  Details follow.

\emph{\underline{Step 1}}: In this step, we use the probabilistic method show that there are~\(1 \times m\) vectors~\(g_1,\ldots,g_k \in \mathbb{F}_q^{m}\) satisfying the following properties:\\
\((1)\) The set of vectors~\({\mathcal G} := \{g_i\}_{1 \leq i \leq k}\) is linearly independent and mutually orthogonal.\\
\((2)\) The~\([m,k]_q-\)space spanned by~\({\mathcal G}\)  has a minimum distance of at least~\(d.\)\\

Letting~\(g_1,\ldots,g_k\) be independent and identically distributed (i.i.d.) vectors in~\(\mathbb{F}_q^{m},\) we show below that properties~\((1)-(2)\) hold with positive probability. We begin a couple of definitions. For~\(2 \leq i \leq k\) let~\(D_i\) be the event that~\({\mathcal G}_i :=\{g_1,\ldots,g_i\}\) is a basis and let~\(E_i\) be the event that the set~\({\mathcal G}_i\) is mutually orthogonal. Also let~\(F_i\) be the event that the space~\({\mathcal C}_i\) spanned by~\({\mathcal G}_i\) has a minimum distance of at least~\(d.\) Finally let~\(J_i := D_i \cap E_i \cap F_i \) and define~\(J_1 = F_1.\)  In what follows, we estimate the conditional probabilities of the events~\(D_i, E_i\) and~\(F_i,\) given~\({\mathcal G}_{i-1},\) in that order.

Given that~\({\mathcal G}_{i-1} = \{g_1,\ldots,g_{i-1}\}\) is a basis, the size of the space spanned by~\({\mathcal G}_{i-1}\) is~\(q^{i-1}.\) Therefore the set~\({\mathcal G}_i\) is a basis if and only if~\(g_i\) is chosen from the remaining~\(q^{m}-q^{i-1}\) vectors and so
\[\mathbb{P}(D_i \mid {\mathcal G}_{i-1}) \ind(D_{i-1}) = \frac{q^{m}-q^{i-1}}{q^{m}} \ind(D_{i-1}) = \left(1-\frac{1}{q^{m-i+1}}\right)\ind(D_{i-1}),\]
where~\(\ind(.)\) refers to the indicator function. Consequently, we get that
\begin{equation}\label{pek}
\mathbb{P}(D_i \mid {\mathcal G}_{i-1}) \ind(J_{i-1}) = \left(1-\frac{1}{q^{m-i+1}}\right)\ind(J_{i-1}).
\end{equation}

Next we look at the conditional probability of the event~\(E_i.\) Given that~\({\mathcal G}_{i-1}\) is a mutually orthogonal basis, we would like to estimate the probability that
\begin{equation}\label{ortho_cond}
g_i \cdot g_j^{T} = 0 \text{ for each } 1 \leq j \leq i-1.
\end{equation}
Because~\(i-1 \leq k \leq m\) (see statement of Lemma) and the event~\(E_{i-1}\) occurs, the~\((i-1) \times m\) matrix
\begin{equation}\label{bi_def}
B_{i-1} := [g^{T}_1,g^{T}_2,\ldots,g^{T}_{i-1}]^{T}
\end{equation}
has a full rank of~\(i-1.\) Moreover, since the matrix~\(B_{i-1}\) is completely determined by the vectors in~\({\mathcal G}_{i-1},\) we assume for simplicity that the first~\(i-1\) columns of~\(B_{i-1}\) are linearly independent. The conditions in~(\ref{ortho_cond}) can then be rewritten as
\begin{equation}\label{g_eq}
B^{(1)}_{i-1} \cdot \left(g^{(1)}_i\right)^{T} = h_{i-1} := -B^{(2)}_{i-1} \cdot \left(g^{(2)}_i\right)^{T},
\end{equation}
where~\(g^{(1)}_i\) is the~\(1 \times (i-1)\) vector formed by the first~\(i-1\) entries of~\(g_i\) and~\(g^{(2)}_i\) is the vector formed by the remaining entries of~\(g_i.\) Similarly,~\(B^{(1)}_{i-1}\) is the~\((i-1) \times (i-1)\) invertible square matrix formed by the first~\(i-1\) columns of~\(B_{i-1}\) and~\(B^{(2)}_{i-1}\) is formed by the remaining columns of~\(B_{i-1}.\) Given~\(g^{(2)}_i\) and~\({\mathcal G}_{i-1},\) we see that~(\ref{g_eq}) holds with probability~\(\frac{1}{q^{i-1}}\) and so averaging over~\(g^{(2)}_i\) we get that
\begin{equation}\label{pfk}
\mathbb{P}(E_i \mid {\mathcal G}_{i-1}) \ind(J_{i-1}) = \frac{1}{q^{i-1}} \cdot \ind(J_{i-1}).
\end{equation}


Finally, we estimate the conditional probability of the event~\(F_{i},\)  that the space~\({\mathcal C}_i\) spanned by the columns of the matrix~\(B_i\) as defined in~(\ref{bi_def}), has a minimum distance at least~\(d.\) Let~\(x = (x_1,\ldots,x_i)\) be any vector in~\(\mathbb{F}_q^{i}\) with~\(x_i \neq 0\) and~\(y\) be any vector in~\(\mathbb{F}_q^{m}.\) Given~\({\mathcal G}_{i-1},\) the conditional probability~\(\mathbb{P}\left(x \cdot B_i = y \mid {\mathcal G}_{i-1}\right) =\frac{1}{q^{m}}\) since the~\(j^{th}\) relation~\(1 \leq j \leq m\) in the matrix equation~\(x \cdot B_i = y \) is of the form \[x_i \cdot g_{i,j} = y_i -\sum_{1 \leq u\leq i-1} x_u \cdot g_{u,j}\] and~\(g_i = (g_{i,1},\ldots,g_{i,m})\) is the random~\(i^{th}\) row of the matrix~\(B_i.\)

In effect, given~\({\mathcal G}_{i-1},\) the probability that the vector~\(x\) maps to \emph{some} vector with weight at most~\(d-1\) is bounded above by~\(\frac{1}{q^{m}} \cdot \sum_{l=0}^{d-1} (q-1)^{l} \cdot {m\choose l}\) and since there are~\((q-1) \cdot q^{i-1}\) choices for~\(x,\) we get the following: Given~\({\mathcal G}_{i-1},\) the probability that \emph{some} vector~\(x\) with~\(x_i \neq 0\) maps to some vector with weight at most~\(d-1\) is bounded above by
\[\frac{(q-1)q^{i-1}}{q^{m}} \cdot \sum_{j=0}^{d-1} (q-1)^{j} \cdot {m\choose j} =: \frac{\theta}{q^{m-i+1}}.\]  By definition, if the event~\(F_{i-1}\) occurs, then no vector~\(x \in \mathbb{F}_q^{i}\) with~\(x_i = 0\) maps to a vector in~\(\mathbb{F}_q^{m}\) with weight at most~\(d-1.\) Summarizing we therefore get that
\begin{equation}\label{pjk}
\mathbb{P}\left(F_i \mid {\mathcal G}_{i-1} \right) \ind(J_{i-1}) \geq \left(1-\frac{\theta}{q^{m-i+1}}\right) \ind(J_{i-1})
\end{equation}

Combining~(\ref{pek}),~(\ref{pfk}) and~(\ref{pjk}) and using~\[\mathbb{P}(A \cap B \cap C) \geq \mathbb{P}(A) - \mathbb{P}\left(B^c\cup C^c\right) \geq \mathbb{P}(A) - \mathbb{P}(B^c) - \mathbb{P}(C^c)\]
with~\(A = E_i\) and~\(B = D_i\) and~\(C = F_i,\) we get
\begin{equation}\label{pdk}
\mathbb{P}(J_i |{\mathcal G}_{i-1}) \ind(J_{i-1}) \geq \left(\frac{1}{q^{i-1}} - \frac{1+\theta}{q^{m-i+1}}\right)\ind(J_{i-1}).
\end{equation}
Taking expectations and using the fact that~\(J_{i} \subset J_{i-1}\) we then get that\\\(\mathbb{P}(J_i) \geq \frac{\epsilon_i}{q^{i-1}} \cdot \mathbb{P}(J_{i-1})\)
where~\(\epsilon_i :=  1 - \frac{1+\theta}{q^{m-2i+2}}.\)
By iteration we therefore get that
\begin{equation}
\mathbb{P}(J_k) \geq \prod_{i=2}^{k} \frac{\epsilon_i}{q^{i-1}} \cdot \mathbb{P}(J_1) = q^{-{k \choose 2}} \cdot \prod_{i=2}^{k} \epsilon_i \cdot \mathbb{P}(F_1) \geq q^{-{k \choose 2}} \cdot \prod_{i=2}^{k} \epsilon_i \cdot \left(1-\frac{\theta}{q^{m}}\right) \nonumber
\end{equation}
since~\(J_1 = F_1\) and~\(\mathbb{P}(F_1) \geq 1-\frac{\theta}{q^{m}}\) by~(\ref{pjk}). Therefore to get~\(\mathbb{P}(J_k) >0,\) it suffices to ensure that~\(\epsilon_k > 0\) which is true if~(\ref{q_cond}) holds. This proves that properties~\((1)-(2)\) hold and thereby completes the proof of Step~\(1.\)


\underline{\emph{Step~\((2)\):}} We begin with the proof of~\((i).\) Suppose~\(q=2^{r}\) for some integer~\(r \geq 1\) and let~\(\beta\) be any primitive element of~\(\mathbb{F}_q.\) The order of~\(\beta\) is~\(q-1\) which is odd and so~\(\alpha := \beta^2\) is also a primitive element.

Letting~\(\{g_1,\ldots,g_k\} \subset \mathbb{F}_q^{m}\) be the vectors satisfying properties~\((1)-(2)\) as obtained in Step~\(1\) above, we define elements~\(\alpha_i \in \mathbb{F}_q, 1 \leq i \leq k\) as follows. For~\(1 \leq i \leq t\) set~\(\alpha_i=0\) if~\(g_i \cdot g_i^{T} =0.\) Else let~\(\omega_i\) be such that~\[\alpha^{\omega_i} = - g_i \cdot g_i^{T}\] and set~\(\alpha_i = \beta^{\omega_i}.\) For~\(t+1 \leq i \leq k,\) we let~\(\alpha_i\) be any element such that~\[\alpha_i^2 + g_i \cdot g_i^{T} \neq 0.\]

Consider the~\([m+k,k]_q-\)space~\({\mathcal C}_1\) spanned by the rows of the matrix
\begin{equation}\label{g_deff}
G_1 := [A\mid B]
\end{equation}
where~\(A\) is the diagonal~\(k \times k\) matrix~\(diag(\alpha_1,\ldots,\alpha_k)\) and~\(B\) is the~\(k \times m\) matrix
\begin{equation}\label{b_def}
B := [g^{T}_1,g^{T}_2,\ldots,g^{T}_{k}]^{T}.
\end{equation}
By construction we get that~\(G_1 \cdot G_1^{T}\)  is a diagonal matrix containing exactly~\(t\) zeros in its diagonal. This implies that~\({\mathcal C}_1 \cap {\mathcal C}_1^{\perp}\) has dimension~\(t.\) To see this is true let~\(h_1,\ldots,h_k\) be the rows of~\(G_1\) and let~\(v \in {\mathcal C}_1 \cap {\mathcal C}_1^{\perp}\) be any vector so that~\(v = \sum_{i=1}^{k} a_i \cdot h_i\) with~\(a_i \in \mathbb{F}_q.\) Taking dot product with~\(h_i\) for~\(t+1 \leq i \leq k\) we get that
\[0 = v \cdot h_i^{T} = a_i \cdot \left(h_i \cdot h_i^{T}\right)\] and since~\(h_i \cdot h_i^{T} \neq 0\) we get that~\(a_i=0\) for~\(t+1 \leq i \leq k.\) This implies that~\(v\) is a linear combination of~\(h_{1},\ldots,h_t\) and therefore proves that~\({\mathcal C}_1 \cap {\mathcal C}_1^{\perp}\) has indeed dimension~\(l.\)

Finally, by property~\((2)\) of the vectors~\(\{g_i\}\) obtained above, the space~\({\mathcal C}_1\) also has a minimum distance of at least~\(d.\) This proves Theorem~\ref{thm_main}\((i).\)

We prove~\((ii)\) as follows.  As before let~\(\{g_1,\ldots,g_k\} \subset \mathbb{F}_q^{m}\) be the vectors obtained in Step~\(1\) and let~\(\Delta\) be a~\(k \times k\) diagonal matrix containing exactly~\(t\) zeros in its diagonal. Because~\(q=p^{r}\) and~\(p \equiv 1 \mod{4}\) there exists an integer~\(0 \leq a \leq p-1\) such that~\(a^{2}\equiv -1 \mod{p}\) (Theorem~\(2.12,\) pp.~\(53,\) Niven et al. (1991)). Therefore letting~\(B\) as in~(\ref{b_def}) and setting
\begin{equation}\label{g_deff2}
G_2 := [\Delta \mid B\mid a B],
\end{equation}
we get that~\(G_2 \cdot G_2^{T} = \Delta^2.\) As in case~\((i),\) the space~\({\mathcal C}_2\) spanned by the rows of~\(G_2\) is a~\([2m+k,k]_q-\)space with hull dimension~\(t\) and moreover, has a minimum distance at least~\(2d\) by construction. This completes the proof of~\((ii).\)

Finally, for any odd prime~\(p,\) there are integers~\(0 \leq a,b \leq p-1\) such that~\(a^{2}+b^2\equiv -1 \mod{p}\) (Theorem~\(5.14,\) pp.~\(246,\) Niven et al. (1991)). We proceed as in case~\((ii)\) and set
\begin{equation}\label{g_deff3}
G_3 := [\Delta \mid B\mid a B \mid b B],
\end{equation}
to get that~\(G_3 \cdot G_3^{T} = \Delta^2.\) Consequently, the space~\({\mathcal C}_3\) spanned  by the rows of~\(G_3\) is a~\([3m+k,k]_q-\)space with hull dimension~\(t\) and minimum distance at least~\(3d.\) This completes the proof of~\((iii)\) and therefore the Theorem.~\(\qed\)

\underline{\emph{Acknowledgements}}: I thank Professors Rahul Roy and C. R. Subramanian for crucial comments and also thank IMSc and IISER Bhopal for my fellowships.

\bibliographystyle{plain}

\end{document}